%% file: ms.tex
\documentclass[preprint]{aastex}
\usepackage{xspace}

\newcommand{\eg}{{\it e.g.}\xspace}

\newcommand{\cf}{{\it cf.}\xspace}
\newcommand{\etal}{{\it et al.}\xspace}
\newcommand{\nifsx}{${}^{56}$Ni\xspace}

\slugcomment{To appear in Ap. J.}

\lefthead{Ivanov \etal}
\righthead{SN~Ia Age-Luminosity Relation}

\begin{document}

\title{On the Relation Between Peak Luminosity and Parent Population
of Type Ia Supernovae: A New Tool for Probing the Ages of Distant 
Galaxies}

\author{Valentin D. Ivanov, Mario Hamuy, Philip A. Pinto}
\affil{Steward Observatory, The University of Arizona, Tucson, AZ 85721,\\ 
	vdivanov, mhamuy, \& ppinto@as.arizona.edu} 

\begin{abstract} 
We study the properties of Type Ia Supernovae (SNe~Ia) as functions of the
radial distance from their host galaxy centers. Using a sample of 62 SNe~Ia
with reliable luminosity, reddening, and decline rate determinations, we
find no significant radial gradients of SNe~Ia peak absolute magnitudes or
decline rates in elliptical+S0 galaxies, suggesting that the diversity of
SN properties is not related to the metallicity of their progenitors.  We
do find that the range in brightness and light curve width of supernovae in
spiral galaxies extends to brighter, broader values. These results are
interpreted as support for an age, but not metallicity, related origin of
the diversity in SNe~Ia.  If confirmed with a larger and more accurate
sample of data, the age-luminosity relation would offer a new and powerful
tool to probe the ages and age gradients of stellar populations in galaxies
at redshift as high as $z\sim1-2$.  The absence of significant radial
gradients in the peak $\rm (B-V)_0$ and $\rm (V-I)_0$ colors of SNe Ia
supports the redding correction method of Phillips \etal (1999).  We find no
radial gradient in residuals from the SN~Ia luminosity-width relation,
suggesting that the relation is not affected by properties of the
progenitor populations and supporting the reliability of cosmological results
based upon the use of SNe~Ia as distance indicators.
\end{abstract} 


\section{Introduction}
In recent years the search for a precise extragalactic distance indicator
has stimulated the study of Type Ia Supernovae (hereafter SN~Ia). A number
of surveys (\eg \citealt{ham93,ham96a,ham96b,ham96c,rie99,per99})
have yielded samples with high-quality light curves and well-defined
selection criteria. The availability of such data provides an opportunity
for studying SNe~Ia at an unprecedented level of detail.

Perhaps the most important result of these surveys is that the SNe~Ia do
not constitute a homogeneous family of objects -- SNe~Ia are not all the
same. They span a range in peak B luminosity of approximately three
magnitudes and exhibit a variety of light curve morphologies
\citep{maz94,ham95}. Their heterogeneity has also been demonstrated
spectroscopically \citep{phi92}.

While the discovery of a significant dispersion in peak brightness seemed 
to spell disaster for the use of SNe~Ia as extragalactic standard candles,
\citet{phi93} demonstrated a tight correlation between the peak luminosity
and the rate of decline from peak in the first 15 days. This relation was
refined by \citet{ham95}, and shown to reduce the dispersion about the
Hubble relation by a factor of two by \citet{ham96b}. This dispersion was
then reduced even further by \citet{phi99}, who
included a precise correction for extinction to achieve a dispersion
between 0.09 and 0.12 magnitudes, depending on photometric band.

\citet{rie96}, using a different statistical technique, have also shown
that the light curve shape near peak is an accurate predictor of luminosity.
\citet{Perlmutteretal97} have demonstrated that not only is there a
correlation of the light curve shape with luminosity but that, near peak,
the light curve is the {\em same} in all supernovae up to a correlated
scaling in time and luminosity. All three methods are capable of achieving
similarly accurate determinations of a supernova's peak brightness.

The discovery and refinement of the luminosity-width relation in SNe~Ia has
enabled \citet{rie99} and \citet{per99} to use observations of SNe~Ia over 
a large range of redshifts to show that the cosmological expansion is 
accelerating at a rate which cannot be explained without the introduction 
of new physics such as a cosmological constant.

It is widely accepted that SNe~Ia result from the thermonuclear
incineration of a carbon/oxygen white dwarf, and that the light curve is
powered by the radioactive decay of \nifsx produced in the explosion.  The
systematic diversity of SNe~Ia may provide important clues to the nature of
their progenitors and the explosion mechanism. It has been suggested that
the range in peak luminosities displayed by SNe~Ia reflects a range in the
nucleosynthetic \nifsx yield. \citet{hof96} demonstrated that a variety of
light curve shapes could be generated from various explosion models, and
that the width of the light curve was correlated in some models with the
\nifsx mass. \citet{PintoE00} showed that changing only the \nifsx yield in
a typical explosion model can reproduce both the slope and the absolute
normalization of the luminosity-width relation to well within the observed
dispersion. Neither of these models explains the origin of SN~Ia diversity,
however.

Unfortunately, much of the most important physics in current models for the
evolution and explosion of SNe~Ia is far too complex to be modeled in
anything close to a predictive fashion. For example, because we lack a
detailed theory of turbulent combustion, the behavior of the nuclear
burning rate, critical to the properties of the resulting explosion, must
be put in ``by hand''. \citet{hof98}, and \citet{ume99a,ume99b} have
explored the effects of progenitor age and metallicity on SN properties
using one particular set of assumptions. They find that in this case the
supernova progenitor's properties affect the $\rm^{56}Ni$ production and
the light curve width, with younger and more metal-rich progenitors leading
to brighter SNe Ia.

An additional route towards understanding the origin of SN~Ia diversity is
the study of the supernova environments, in particular the location of the
supernov\ae\ in the host galaxy. Since properties of the stellar population
differ from galaxy to galaxy and vary with location within a galaxy, such
studies can show which properties of the parent population affect the
nature of the SN~Ia explosion.

\citet{ham96a} found that faster-declining (fainter)
SNe~Ia tend to occur in earlier-type galaxies, suggesting that the age of
the progenitor is a significant parameter. This result was confirmed by
\citet{rie99}, who also found that after correcting
peak absolute magnitudes for extinction and applying the width-luminosity
relation, there was no residual difference in the corrected supernova
magnitudes between early- and late-type galaxies.

The location of SNe in their host galaxies has been extensively studied
(\cf \citealt{van97} and references therein). \citet{sha79} noted a
deficiency of SNe~Ia in the innermost regions of galaxies, but attributed
it to a selection effect; in photographic surveys, it is difficult to
discern SNe against the higher background on which they are projected near
the centers of the galaxies. \citet{wan97} and \citet{how00} later confirmed this effect, at least in the older photographic SNe searches. \citet{ham99}
also showed this to be the case in a more rigorous analysis of the
Cal\'an-Tololo survey.

Major problems which have plagued these and similar studies of the radial
gradients of SNe~Ia in their host galaxies include the difficulty of
determining true galactocentric distances and the intrinsic spread of
properties among the host galaxies themselves.  The goal of this paper is
to use the best-observed sample of SNe~Ia available to study the radial
distribution of SNe~Ia in their host galaxies and the variations of their
properties with galactocentric distance. We improve on these previous
studies by: (i) deprojecting the observed angular distances between the SN
and the host galaxy center, removing ambiguities arising from projecting 
the distances on the plane of the sky; (ii) normalizing such
distances to the sizes of the hosts; (iii) applying statistical methods to
estimate the significance of radial gradients in SN properties; and (iv)
exploring the possible applications of SNe~Ia as tools to probe the stellar
populations of distant galaxies. We will compare our results with current
theoretical understanding of the origin of the peak luminosity - decline
rate relation, and will assess consistency with the known radial variation
of age and metallicity in galaxies.

\section{The SNe~Ia Sample: Selection and Properties}

Our sample consists of 62 well-observed SNe~Ia (listed in
Table~\ref{tbl-1}) with reliable estimates of peak luminosity, reddening,
and decline rate $\rm\Delta m_{15}(B)$ as determined by \citet{phi99};
throughout this paper we use only the reddening-corrected absolute
magnitudes as presented in that work. Several SNe with no peak BVI
measurements were used exclusively in tests that require only positional
information.  SN1992K and SN1991bg were excluded from our analysis as
peculiarly subluminous; there is considerable debate in the supernova
community as to whether these objects are a different sort of explosion or
merely the faintest tail of the ``normal'' supernovae distribution.

The Hubble type and redshifts of the host galaxies were taken from the
NASA/IPAC Extragalactic Database (hereafter NED). The galaxy semi-major
axes and position angles are taken from the RC3 \citep{dev91}, the UGC
\citep{nil73}, and the ESO/Upsala catalogs \citep{hol74}. For galaxies with
no available measurements in the literature, we determined galaxy
parameters from the Cal\'an/Tololo B-band images. SN peak magnitudes and
decline rates were determined by template-fitting following the
prescription of \citet{ham96a}. We use Cepheid distances measured by
\citet{gib00} for six nearby SNe, and distance derived from observed
redshifts for the remainder of the sample. Throughout this paper we adopt
a value for the Hubble constant $\rm H_0~=~69~km~sec^{-1}~Mpc^{-1}$ for
consistency with the Gibson \etal distances. We also assume a cosmology
with $\rm \Omega_0=0.2$, $\rm \Omega_\Lambda=0.0$ to calculate the linear 
galactocentric distances.

\placetable{tbl-1}

To de-project the angular distances of SNe Ia from the centers of disk
galaxies, we assume that all SNe~Ia populate disks which are thin compared
with the galaxy diameters. This is a reasonable assumption for spirals and,
perhaps, for S0 galaxies. To compute actual galactocentric distances, we
express the SN position in a radial coordinate system centered at the host
galaxy nucleus and aligned with the semi-major axis. We then use the axial
ratio to calculate the inclination of the disk. In most cases the projected
angular offsets (listed in Table~\ref{tbl-1}) come from the discovery
announcements of the supernovae, with positions verified against the
NED. We did not de-project the angular distances for the SNe in ellipticals
as the depths of these galaxies along the line of sight are comparable to
their apparent sizes. 

\begin{figure}[!t]
\epsscale{0.8}
\plotone{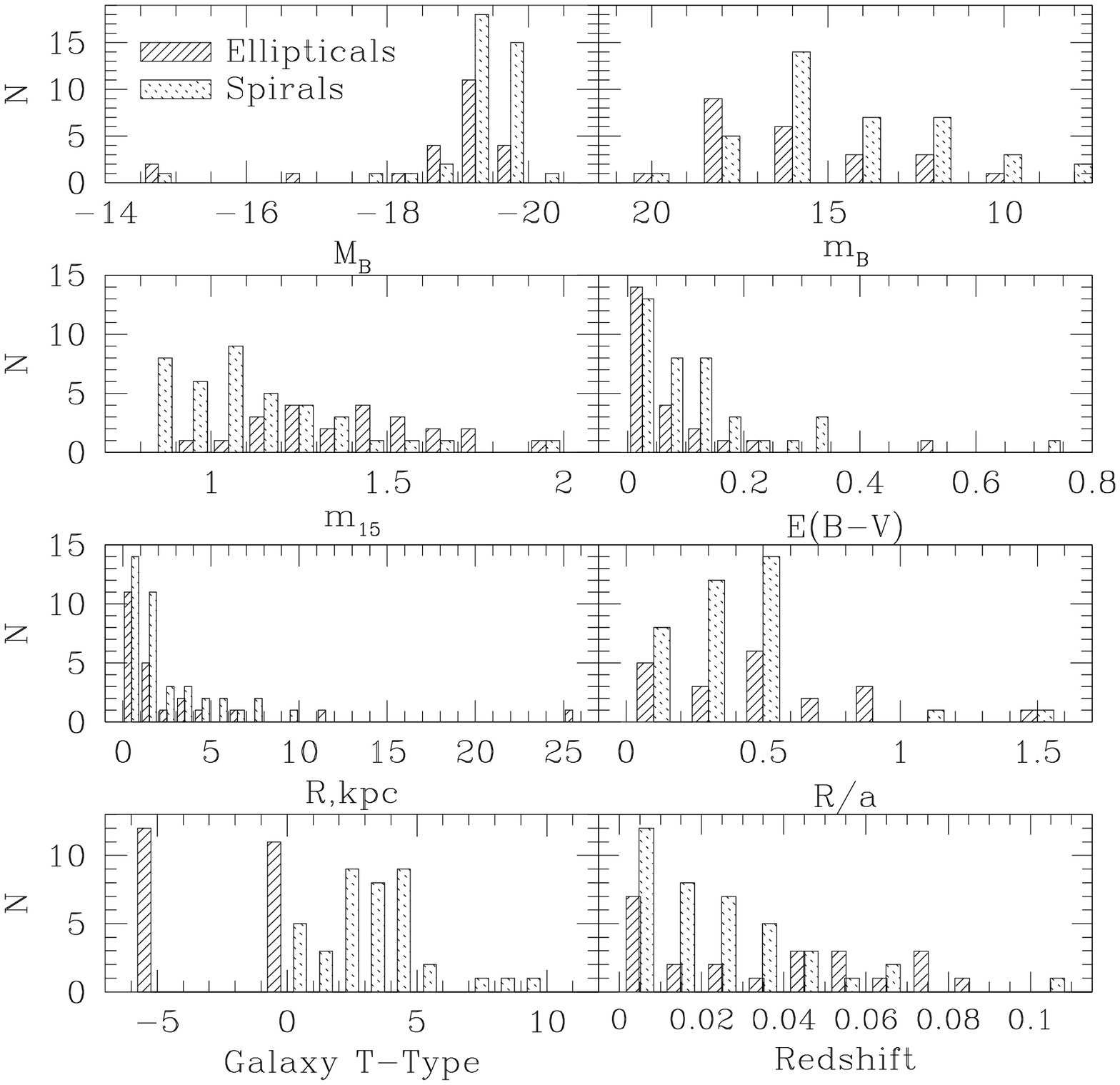}
\caption{Properties of the SNe~Ia sample. The distributions of SNe Ia in
elliptical and S0 galaxies are given with solid lines, and the
distributions of the SNe Ia in spirals are given with dashed lines.}
\label{fig1}
\end{figure}

\placefigure{fig1}

The distributions of the major properties of our sample are shown in
Figure~\ref{fig1}. 23 SNe are in elliptical and S0 hosts and 39 in spirals.
Observational selection \citep{ham99} accounts for
the higher redshift of the SNe in elliptical+S0 than in spirals; early-type
galaxies tend to dominate the statistics for larger projected
galactocentric distances. Reddening in spirals is seen to be higher than in
early-type galaxies, with the notable exception of SN1986G ($\rm
E(B-V)=0.50\pm0.05$ mag) whose host is the peculiar S0 Seyfert~2 galaxy
NGC~5128. Naturally, we find in our sample a deficiency of SNe fainter
than $\rm M_B = -19$ mag, with $\rm E(B-V)>0.10$ mag.

\placetable{tbl-2}

The brightness at maximum light of SNe~Ia in our sample is related to the
morphological type of the host galaxy. \citet{ham96a} pointed out that
faster-declining (fainter) SNe tend to occur in earlier type galaxies, a
result later confirmed by \citet{rie99}. We carried out a
Kolmogorov-Smirnov (hereafter KS) test between the peak absolute magnitude
distributions for the SNe in early-type and late-type galaxies. The
significance level P for the null hypothesis that the two sets of data are
drawn from the same parent distribution is given in
Table~\ref{tbl-2}. Small values of P suggest that the distributions are
different. We obtained only 2, 2 and 7\% of probability, that the
distributions of SNe Ia peak magnitudes, respectively for B, V, and I-band
observations, in elliptical+S0 galaxies and spirals were drawn from the
same parent distribution. The higher probability for the I-band reflects
the smaller amplitude of the luminosity-width relation in the near infrared. 
The rate of decline is an even better indicator of a difference between SNe~Ia 
in early- and in late-type galaxies, giving a significance level of only
0.001\%.  The de-reddened $\rm (B-V)_0$ SNe colors, on the other hand, are
identical to a confidence level of 90\%. The very small uncertainties in
this analysis reflect major differences of the intrinsic properties of the
host galaxies such as age, metallicity, and dust content, and strongly
suggest that the properties of SNe~Ia are determined by their environments.

\section{Radial Gradients of SNe Ia Properties}

The SN host galaxies show a spread of sizes, masses and metallicities --
presumably because larger and more massive galaxies can better retain the
heavy elements produced by stellar evolution
\citep{ari87}. Galaxy-to-galaxy abundance variations can be corrected, to
some extent, by a proper normalization of galactocentric
distances. Following \citet{van97}, we divided the
galactocentric distances by the semi-major axis. This approach has the
advantage that the semi-major axis can be used as a "metalometer". Since
semi-major axes are not always accurately known, we will carry out our
analyses using both normalized and absolute radial distances, for
comparison.

\begin{figure}[!t]
\epsscale{0.8}
\plotone{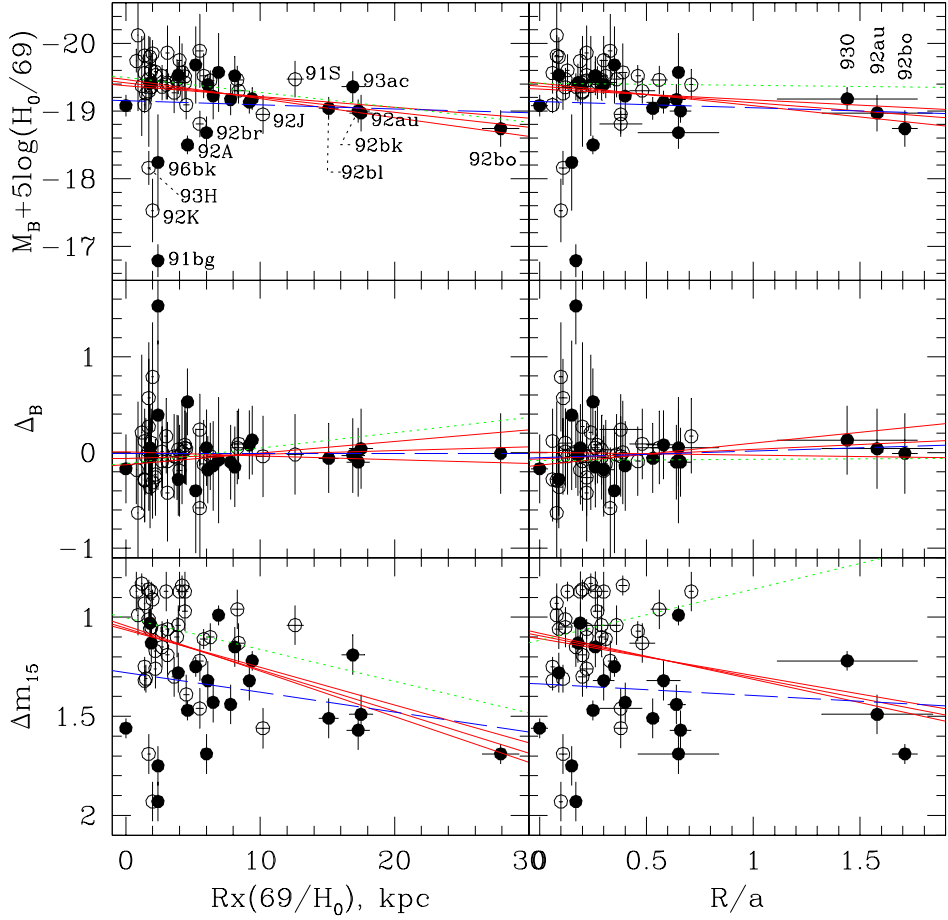}
\caption{SN parameters as function of the galactocentric distances (left
panels), and galactocentric distance normalized to the semi-major axis
(right panels). The absolute B-band magnitude $\rm (M_B)$ is shown at the
top. The residuals from the peak luminosity vs. decline rate relation
$(\Delta_B)$ are shown at the center. The decline rate $\rm (\Delta m_{15})$ 
is shown at the bottom. The open circles represent SNe Ia in spirals and the
solid dots represent SNe Ia in elliptical and S0 galaxies. The fit to the
entire sample and its $1\sigma$ errors are shown with solid lines, the fit
to the SNe Ia in elliptical and S0 galaxies is shown with a dashed line and
the fit to the SNe Ia in spirals is shown with a dotted line.}
\label{fig2}
\end{figure}

\placefigure{fig2}

We first carried out linear fits for the absolute peak magnitude against
the linear distance $\rm R_{dp}$ (in Kpc) and for the distance normalized
to the semi-major axis $\rm R_{dp}/a$. We then calculated the residuals
$\Delta$ from the luminosity-width relation for each SN, using the
quadratic form of the relation given by \citet{phi99}, in order to look for
trends of $\Delta$ with galactocentric distance. We then computed fits for
the decline rate $\rm \Delta m_{15}$ against both
distances. Table~\ref{tbl-3} lists the coefficients, their $1\sigma$ errors
in brackets, and the standard deviation of the fits shown in
Figure~\ref{fig2}. We carried out separate analyses for the entire sample,
for SNe in elliptical+S0 galaxies, and for SNe in spirals. We concentrate
on fits to B-band magnitudes and widths because the amplitude of the
luminosity-width rate relation is the largest at these wavelengths,
rendering our results both more sensitive to differences among supernovae
and to the de-reddening procedure employed by Phillips \etal.

\placetable{tbl-3}

Taking into account the entire sample, we find that SNe closer to the
galaxy centers are brighter, in agreement with the conclusion of
\citet{wan97}. However, when we divide the SNe Ia according to the host
galaxy type, we find no significant radial gradient of the absolute peak
magnitudes. The gradient in the sample as a whole can easily be understood;
a brighter population of SNe Ia in spirals dominates the sample at small
galactocentric distances $\rm (R_{dp} \leq 7.5)$ while the fainter
population of SNe Ia in ellipticals and S0's dominates at larger radial
distances $\rm (R_{dp} > 7.5)$.

There seems to be a small but significant $(2-5\sigma)$ correlation between
decline rate and radial distance, but this result is due mainly to one
object (SN1992bo) illustrating the importance of extending the SN sample.
When the early- and late-type galaxies are treated separately, the decline
rate gradient is reduced by a factor of 2 or more. Employing normalized
radial distances makes the trends of $\rm \Delta m_{15}$ with distance
disappear at the $2-2.5\sigma$ level.

\begin{figure}[!t]
\epsscale{0.8}
\plotone{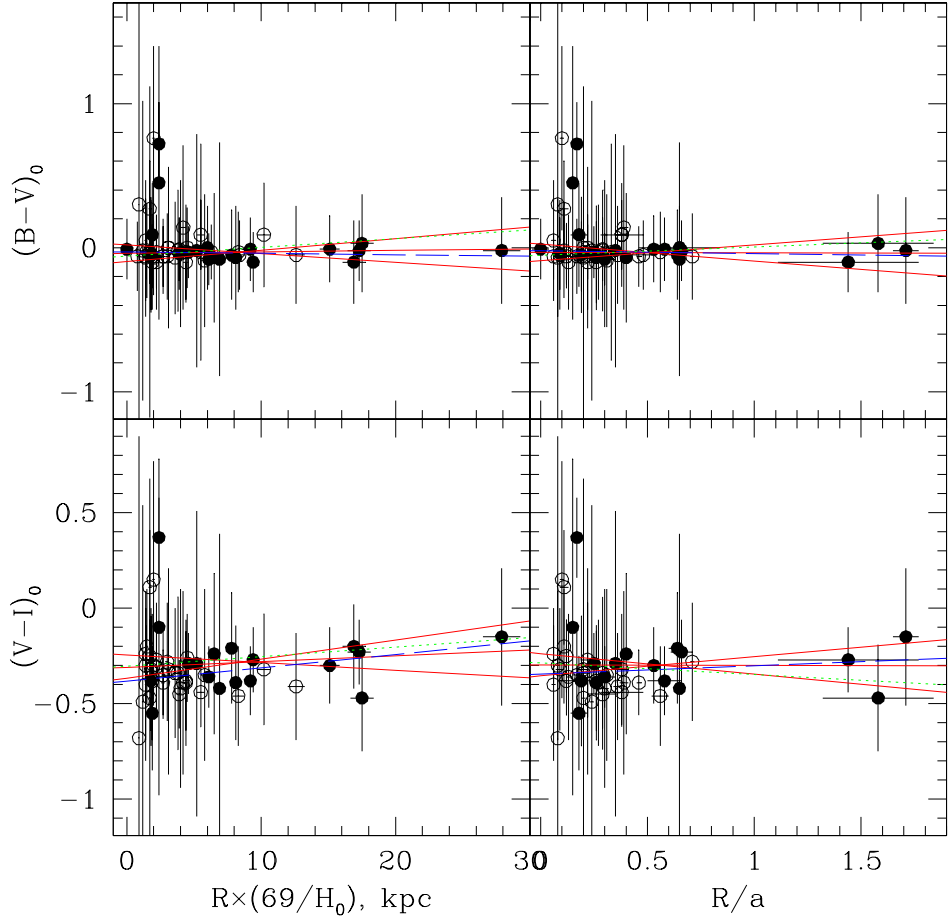}
\caption{
The reddening-corrected SN colors as function of 
the galactocentric distances (left panels), and galactocentric distance 
normalized to the semi-major axis (right panels). $\rm (B-V)_0$ color is 
shown at the top panel, and $\rm (V-I)_0$ color is shown at the bottom 
panel. The fits are indicated as in Figure 2. 
}
\label{fig3}
\end{figure}

\placefigure{fig3}

The de-reddened SNe~Ia $\rm (B-V)_0$ color at maximum light shows no radial
variation (Figure~\ref{fig3}). If we use the linear distance fits, the
color change is only 0.03 mag over 30 Kpc for the entire sample. The change
is just sightly larger for spirals alone (0.06 mag over 10 Kpc). Similarly,
there is no radial gradient of $\rm (V-I)_0$ color at peak within the
uncertainties. This results lend credence to the reddening correction
method of Phillips \etal.

\section{Discussion} 

The radial metallicity gradient in elliptical galaxies is an established
observational fact (\citealt{hen99} and references therein;
\citealt{kob99,tam00}). At the same time, the population in these galaxies
is co-eval \citep{wor94,vaz96,vaz97}.  The stars near the center of spiral
galaxies are, however, in general both more metal rich {\em and} older when
compared with outer regions \citep{hen99,bel00}. This difference provides
an ideal opportunity to separate the effects of age and metallicity on
SNe~Ia properties.

The lack of any radial gradients in SNe~Ia properties in elliptical+S0
galaxies suggests that neither the absolute peak magnitude nor the decline
rate is a strong function of the metal abundance of the progenitor.  The
residuals from the luminosity-width relation seem also to be independent of
metallicity in the host population. At the same time, SNe~Ia in late-type
galaxies exhibit a significantly different distribution in both peak
magnitude and decline rate from their cousins in early-type galaxies.

These results are consistent with the hypothesis that the {\em age} of the
progenitor, but not the metallicity, may be responsible for the diversity
of SNe Ia peak luminosities and decline rates.  Current theoretical
understanding of the problem is fraught with uncertainties, but a possible
scenario was suggested by \citet{ume99a,ume99b}. They argue that, no matter
what the details of explosive burning may be, the larger specific energy
available from burning larger mass fractions of carbon in a C/O dwarf will
result in greater \nifsx production and, hence, brighter SNe~Ia. They then
argue that the oldest SN progenitor systems are those with the
smallest-mass companions, and that these systems can transfer less mass in
all to the pre-SN white dwarf.  Thus, the pre-SN systems which take longest
to evolve to explosion must be those with the largest initial C/O white
dwarf mass. Since larger C/O core masses have smaller carbon abundances,
these older systems will lead to lower \nifsx masses, and eventually,
fainter supernovae.

Such a scenario naturally explains the absence of bright SNe~Ia in
ellipticals, with their older populations, and is also consistent with 
the larger dispersion of peak luminosity found near the center of spiral
galaxies \citep{wan97} where SNe Ia would be produced from a mixture of
stellar populations with various ages. 

Our results are consistent with the predictions of \citet{von97} who 
carried out a semi-empirical modeling of the white dwarf mass function, 
and derived the SN Ia luminosity function for a range of progenitor 
masses and ages. Their work used the older sub-Chandrasekhar explosion 
models of \citet{woo94}, different from the Chandrasekhar mass 
explosions considered by \citet{ume99a,ume99b}. 

The range in accretion rate necessary to produce SNe~Ia from Chandrasekhar
white dwarfs in single-degenerate systems is quite narrow \citep{NK91}, too
narrow to lead to a SNe~Ia rate in accord with observations.
\citet{HKN96,HKN99} have suggested that in systems with greater mass
transfer rates, a wind may remove all but the necessary amount from the
system, leading to a much broader range in binary systems capable of
leading to SNe~Ia.  The wind is driven by radiation pressure on metal
lines, and its strength thus depends upon both the metallicity of the
system and the luminosity from burning accreted material. This luminosity
is larger for larger C/O dwarf masses, and thus for a given C/O dwarf mass
the metallicity of the system must be greater than some minimum value.
Below a critical metallicity, the radiation momentum absorbed by the wind
is insufficient to reduce the mass transfer rate below that leading to a
common-envelope system. The papers by Umeda \etal suggest that this will
lead to a metallicity effect on the range of SN~Ia luminosities -- the
larger the C/O dwarf mass, the lower the critical metallicity required for
the wind. Thus, lower-metallicity systems will tend to have larger initial
C/O dwarf masses and hence fainter supernov\ae.

This metallicity effect seems not to be supported by our results. Since 
we do not find strong radial variation of the SNe Ia parameters separately 
in elliptical+S0 galaxies, and in spirals, such an
effect seems unlikely to explain the lack of bright supernovae in the
ellipticals. Indeed, as we find no variation in supernova properties with
radius in early-type galaxies, any wind in SN~Ia progenitor systems must be
much less sensitive to metallicity than posited by Hachisu, Kato, \& Nomoto.

Given a range in \nifsx yield, \citet{PintoE00} show that this is in itself
sufficient to lead to the observed luminosity-width relation. They suggest
that this relation is thus unlikely to be affected by metallicity effects
on the radiation transport responsible for the relation, and the lack of
any radial gradient in SN properties in ellipticals and S0's bears this
out.

Most of the SNe Ia with Cepheid calibrations \citep{gib00} happen to lie at
large radial distances in their host galaxies and are therefore fainter
than average. One might be concerned that a selection bias against SNe~Ia
at small angular separations from their host galaxy centers (the ``Shaw
effect'') would make local and distant samples of SNe Ia incommensurate.
Our results imply that after correcting for extinction and luminosity-width
relation, SNe Ia become true standard candles, independently of their
galactocentric distance, and thus, metal abundance.

Indeed, the lack of any correlation found to the {\em corrected} supernovae
magnitudes should be heartening to those employing SNe~Ia at large
redshifts to measure cosmological properties. The range in both metallicity
and age spanned by the galaxies in our sample exceeds the difference in
average values of these parameters between the local universe and at a
redshift $z\sim1$. To explain away the cosmological results found by 
\citet{Riessetal98} and by \citet{per99}, the corrected peak magnitudes 
of the lower metallicity and/or younger supernovae found at $z\sim1$ 
would have to be fainter on average by more than 0.2 mag, a result 
clearly ruled out by  Figure~(\ref{fig2}).

At lesser distances, the lack of any gradient in residuals from the
luminosity-width relation argues against a bias in the determination of 
the Hubble constant based on intrinsically fainter nearby SNe Ia.

A possible test of the relation between the peak luminosities (and light 
curve shapes) of SNe~Ia and the age of their progenitor populations can be 
carried via spectral indices age diagnostics of the host galaxy population 
\citep{wor94,vaz96,vaz97}. If confirmed, this relation might be used to 
probe the ages and age gradients of stellar populations in galaxies. The
intrinsic brightness of these SNe makes it possible to carry out such an
analysis to redshifts beyond $z\sim1$ with current observational
capabilities. Such a technique is rendered even more important by the
advent of new telescopes and instruments in the near future. For instance,
\citet{dah99} estimate that NGST is expected
to detect about 45 SNe Ia per field in a single $10^4$ sec exposure, with
mean redshift of $\rm z\approx2$. From the ground, the 8.4m wide-field Dark
Matter Telescope proposed by \citet{hin99} would provide high-quality 
observations of many hundreds of supernovae per year to $z\sim1.5$ (Pinto 
1999, private communication).

\section{Summary}

Our KS tests confirm the result of \citet{ham96a}
that SNe~Ia in elliptical+S0 and in spiral host galaxies are intrinsically
different, with faster-declining (fainter) SNe~Ia occurring mostly in
earlier-type galaxies.

We find no significant radial variation of the SNe~Ia peak absolute
brightness in ellipticals+S0s and in spirals. Taken together, SNe do show a
correlation in peak brightness with radial distance. This is a consequence
of the fact that fainter SNe~Ia are found in ellipticals, and these SNe
dominate the sample at larger galactocentric distances.

Our analysis shows no significant radial gradient of the
reddening-corrected peak magnitude $\rm (B-V)_0$ and $\rm (V-I)_0$ colors,
lending support to the reddening correction method of \citet{phi99}.

The peak absolute luminosity and the decline rate of SNe Ia are 
independent of the metallicity of the SNe progenitors, within the 
observational uncertainties. Determinations of extragalactic 
distances and related cosmological parameters are thus not biased by 
metallicity- or age-related evolutionary effects. 

Our results are consistent with an age-related origin of the
luminosity-width relation. If confirmed, this age {\em vs.} decline rate
relation in SNe Ia offers an important new tool to probe the ages and age
gradients of stellar populations in galaxies at redshifts as high as
$z\sim1-2$.

\acknowledgments

We are grateful to Dr. M.M. Phillips for the useful discussions.  The
research has made use of the NASA/IPAC Extragalactic Database which is
operated by the Jet Propulsion Laboratory, California Institute of
Technology, under contract with NASA.  This work has been supported by the
National Science Foundation (CAREER grant AST9501634). PAP
gratefully acknowledges support from the Research Corporation though a
Cottrell Scholarship.

\clearpage

\input{tab1}
\clearpage
\input{tab2}
\clearpage
\input{tab3}

\clearpage

\end{document}

%% file: tab1.tex
\begin{deluxetable}{lllrcccccccccc}
\tabletypesize{\tiny}
\rotate
\tablenum{1}
\tablewidth{0pt}
\tablecaption{Data for the supernovae sample.\label{tbl-1}}
\tiny
\tablehead{
\multicolumn{1}{c}{SN}& 
\multicolumn{1}{c}{Host Galaxy}&
\multicolumn{1}{c}{Hubble}&
\multicolumn{1}{c}{T}&
\multicolumn{1}{c}{Redshift}&
\multicolumn{1}{c}{$\rm a\times b$}& 
\multicolumn{1}{c}{P.A.}& 
\multicolumn{1}{c}{$\rm M_B/ \sigma$}& 
\multicolumn{1}{c}{$\rm M_V/ \sigma$}& 
\multicolumn{1}{c}{$\rm M_I/ \sigma$}& 
\multicolumn{1}{c}{$\rm E(B-V)/\sigma$}& 
\multicolumn{1}{c}{$\rm \Delta m_{15}/ \sigma$}& 
\multicolumn{1}{c}{$\rm R_{p}$}& 
\multicolumn{1}{c}{$\rm R_{dp}$}\\
\multicolumn{1}{c}{19xx}& 
\multicolumn{1}{c}{Name}&
\multicolumn{1}{c}{Type}&
\multicolumn{1}{c}{Type}&
\multicolumn{1}{c}{$\rm z^{helio}$}&
\multicolumn{1}{c}{arcmin}& 
\multicolumn{1}{c}{deg}& 
\multicolumn{1}{c}{mag}& 
\multicolumn{1}{c}{mag}& 
\multicolumn{1}{c}{mag}& 
\multicolumn{1}{c}{mag}& 
\multicolumn{1}{c}{mag}& 
\multicolumn{2}{c}{arcsec}
}
\startdata
 90O&         MCG+03-44-003&              SBa& 1&0.030664& 0.80x0.80&125&-19.46/21&-19.43/19&-18.97/18&0.02/03&0.96/10& 27& 27\\
 90T&              PGC63925&          SA(s)00& 0&0.040400& 1.30x0.70& 85&-19.52/28&-19.45/23&-19.06/20&0.09/04&1.15/10& 19& 20\\
 90Y&              FCCB1147&          E(M32?)&-5&0.039200& 0.46x0.46&  0&-19.42/28&-19.51/23&-18.96/20&0.23/04&1.13/10&  5&  5\\
90af&  [TB93]2131.14-6257.7&              SB0& 0&0.050300& 0.47x0.42&  0&-19.08/16&-19.07/13&     -   &0.04/03&1.56/05&  0&  0\\
 91S&              UGC05691&               Sb& 3&0.055000& 1.10x0.80&140&-19.47/27&-19.42/21&-19.01/19&0.06/04&1.04/10& 18& 24\\
 91U&                IC4232&          Sbc:pec& 4&0.031442& 1.10x0.30&  2&-19.80/29&-19.75/24&-19.43/21&0.11/04&1.06/10&  3&  6\\
91ag&                IC4919&       SB(s)dmPec& 8&0.014223& 1.50x0.70& 31&-19.75/36&-19.71/36&-19.29/37&0.07/03&0.87/10& 21& 27\\
 92J&   [M92b]100643-2624.0&                S& 5&0.045000& 1.00x0.70& 85&-18.95/28&-19.04/22&-18.72/19&0.03/04&1.56/10& 18& 23\\
 92K&           ESO269-G057&      (RS)SAB(r)b& 3&0.010274& 3.20x2.30& 54&-17.53/47&-18.29/44&-18.44/43&0.00/04&1.93/10& 14& 19\\
 92P&                IC3690&              Sb:& 3&0.025401& 1.10x0.30&  6&-19.51/21&-19.41/19&-19.03/18&0.07/03&0.87/10& 10& 17\\
92ae& [WM92]212426.8-614612&              E1?&-5&0.070000& 1.10x0.90&  0&-19.53/20&-19.49/15&     -   &0.12/04&1.28/10&  5&  5\\
92ag&           ESO508-G067&               S?& 5&0.025164& 1.00x0.20&101&-19.42/25&-19.42/22&-19.10/19&0.10/04&1.19/10&  6& 12\\
92al&           ESO234-G069&          SAB(s)c& 5&0.014600& 2.10x0.70&111&-19.52/33&-19.43/32&-19.08/32&0.01/03&1.11/05& 23& 39\\
92aq&   [HM92]230149-3736.8&              Sa?& 1&0.101000& 0.26x0.21&  0&-18.81/19&-18.90/14&-18.46/12&0.00/04&1.46/10&  6&  6\\
92au&  [HM92a]000808-5013.6&               E1&-5&0.062000& 0.31x0.26&  0&-18.97/27&-19.00/21&-18.53/18&0.00/04&1.49/10& 23& 23\\
92bc&           ESO300-G009&              Sab& 2&0.020160& 1.20x0.40& 44&-19.60/24&-19.50/23&-19.14/23&0.00/02&0.87/05&  5&  9\\
92bg&[MH92d]074119.0-622406&               Sa& 1&0.035000& 0.76x0.29& 40&-19.36/20&-19.30/17&-18.95/15&0.01/03&1.15/10&  6&  8\\
92bh&[MH92e]045840.0-585410&              Sbc& 4&0.044000& 0.63x0.37&135&-19.35/17&-19.29/14&-18.91/12&0.12/03&1.05/10&  3&  4\\
92bk&ESO156-G008           &           S0-pec& 0&0.059007& 0.77x0.45&  0&-19.00/17&-18.98/13&-18.75/11&0.01/03&1.57/10& 70& 81\\
92bl&           ESO291-G011&     (R1)SB(s)0/a& 0&0.044000& 1.10x0.80&108&-19.04/17&-19.03/15&-18.73/13&0.00/03&1.51/10& 25& 35\\
92bo&           ESO352-G057&       SB(s)00pec& 0&0.018963& 1.40x0.40& 12&-18.74/27&-18.72/26&-18.57/25&0.00/03&1.69/05& 74&144\\
92bp& [M92n]033422.3-183104&            E2/S0&-5&0.080000& 0.45x0.33& 45&-19.39/14&-19.31/12&-18.95/10&0.00/03&1.32/10&  6&  6\\
92br& [MH93]014355.4-562057&               E0&-5&0.088000& 0.19x0.18&  0&-18.68/24&-18.68/16&     -   &0.01/04&1.69/10& 14& 14\\
92bs&              FCCB0602&          Sc(s)II& 5&0.064000& 0.48x0.32&120&-19.30/19&-19.25/15&     -   &0.10/04&1.13/10& 23& 27\\
 93B&[MH93a]103235.1-341103&              SBb& 3&0.069000& 0.34x0.24&140&-19.42/17&-19.41/14&-18.96/12&0.12/03&1.04/10&  5&  6\\
 93H&           ESO445-G066&          SB(r)ab& 2&0.024247& 1.10x1.00&  0&-18.16/25&-18.43/22&-18.54/20&0.05/04&1.69/10&  7&  7\\
 93O&   [HM93]132819-3257.6&           E5/S01&-5&0.051000& 0.22x0.09& 80&-19.18/16&-19.08/13&-18.81/11&0.00/03&1.22/05& 14& 14\\
93ag&  [HM93a]100125-3513.1&           E3/S01&-5&0.070000& 0.40x0.24&110&-19.13/17&-19.12/14&-18.74/12&0.07/03&1.32/10& 13& 13\\
93ah&           ESO471-G027&          SBb(rs)& 3&0.029494& 1.00x0.40&  2&-19.27/30&-19.20/25&-18.86/23&0.01/04&1.30/10& 12& 12\\
93ac&           CGCG307-023&                E&-5&0.049304&     -    &130&-19.36/23&-19.26/18&-19.06/13&0.12/04&1.19/10& 35& 35\\
93ae&               UGC1071&                E&-5&0.019049& 1.40x1.20&  0&-19.22/34&-19.15/30&-18.91/29&0.00/03&1.43/10& 29& 29\\
 94M&               NGC4493&                E&-5&0.022955& 0.86x0.67&148&-19.17/23&-19.12/21&-18.91/20&0.08/03&1.44/10& 28& 28\\
 94Q&             PGC059076&               S0& 0&0.028965& 0.54x0.31& 90&-19.41/31&-19.36/25&-18.98/23&0.06/04&1.03/10&  4&  6\\
 94S&               NGC4495&              Sbc& 4&0.015166& 1.40x0.80&130&-19.46/30&-19.44/29&-19.10/28&0.00/03&1.10/10& 16& 24\\
 94T&           CGCG016-058&               Sa& 1&0.034664&     -    & 55&-19.09/21&-19.09/18&-18.83/15&0.09/04&1.39/10& 13& 13\\
94ae&               NGC3370&               Sc& 5&0.004265& 3.20x1.80&148&-19.29/82&-19.26/81&-18.79/81&0.12/03&0.86/05& 26& 38\\
 95D&               NGC2962&               S0& 0&0.006545& 2.60x1.90&  3&-19.57/58&-19.49/57&-19.07/57&0.04/02&0.99/05& 94&101\\
 95E&               NGC2441&               Sb& 3&0.011558& 2.00x1.70& 20&-19.86/40&-19.86/39&-19.53/38&0.74/03&1.06/05& 22& 26\\
95ac&    P95cJ224541-0845.2&                S& 5&0.049990&     -    & 55&-19.85/20&-19.78/16&-19.48/13&0.08/04&0.91/05&  4&  4\\
95ak&                IC1844&              Sbc& 4&0.023008& 0.70x0.25&105&-19.53/24&-19.43/22&-19.16/21&0.18/03&1.26/10&  8&  9\\
95al&               NGC3021&        SA(rs)bc:& 4&0.005139& 1.60x0.90&110&-19.37/74&-19.35/73&-18.86/73&0.15/03&0.83/05& 17& 23\\
95bd&               UGC3151&                S& 5&0.015991& 1.10x0.25& 97&-19.57/43&-19.71/37&-19.32/31&0.15/06&0.84/05& 23& 26\\
 96C&          MCG+08-25-47&               Sa& 1&0.029572& 1.00x0.50& 10&-19.42/24&-19.39/22&-19.00/25&0.09/03&0.97/10&143&200\\
 96X&               NGC5061&               E0&-5&0.006775& 3.50x3.00&  0&-19.68/57&-19.66/57&-19.37/56&0.01/02&1.25/05& 60& 60\\
 96Z&               NGC2935&               Sb& 3&0.007584& 3.60x2.80&  0&-19.89/54&-19.86/52&-32.88/50&0.33/04&1.22/10& 69& 70\\
96ai&               NGC5005&             SBcd& 6&0.003169& 5.80x2.80& 65&-20.12/112&-20.42/112&-19.74/111&1.44/04&0.99/10& 22& 28\\
96bk&               NGC5308&               S0& 0&0.006806& 3.70x0.70& 60&-18.24/71&-18.69/63&-18.59/62&0.19/05&1.75/10& 24& 34\\
96bl&[P96]J003618.17+112334&              SBc& 5&0.035965&     -    &  0&-19.58/19&-19.55/17&-19.24/15&0.08/03&1.17/10&  6&  6\\
96bo&                NGC673&               Sc& 5&0.017254& 2.10x1.70&  0&-19.56/30&-19.61/29&-19.21/28&0.28/03&1.25/05&  6&  8\\
96bv&               UGC3432&             Scd:& 6&0.016706& 1.70x0.40&136&-19.82/30&-19.75/28&-19.45/27&0.21/03&0.93/10&  6&  8\\
 37C&                IC4182&           SA(s)m& 9&0.001071& 6.00x5.50&  0&-19.74/17&-19.68/17&     -   &0.03/03&0.87/10& 61& 68\\
 72E&               NGC5253&         ImpecHII&10&0.001348& 5.00x1.90& 45&-19.39/22&-19.33/21&-19.05/23&0.01/03&0.87/10&119&213\\
 80N&               NGC1316& (R)SAB(s)00LINER& 0&0.005871&12.00x8.50& 50&      -  &      -  &      -  &0.05/02&1.28/04&214&270\\
 81B&               NGC4536&     SAB(rs)bcHII& 4&0.006031& 7.60x3.20&130&-19.45/15&-19.42/12&      -  &0.11/03&1.10/07& 56& 99\\
 86G&               NGC5128&         S0pecSy2& 0&0.001825&25.70x20.00& 35&      -  &      -  &      -  &0.50/05&1.73/07&106&126\\
 89B&               NGC3627&        SAB(s)bSy& 3&0.002425& 9.10x4.20&173&-19.26/24&-19.24/22&-19.04/20&0.34/04&1.31/07& 51& 59\\
 90N&               NGC4639&   SAB(rs)bcSy1.8& 4&0.003369& 2.80x1.90&123&-19.52/16&-19.46/13&-19.07/11&0.09/03&1.07/05& 62& 77\\
 91T&               NGC4527&SAB(s)bcHII/LINER& 4&0.005791& 6.20x2.10& 67&      -  &      -  &      -  &0.14/05&0.94/05& 51& 83\\
91bg&               NGC4374&               E1&-5&0.003336& 6.50x5.60&135&-16.79/24&-17.51/17&-17.88/12&0.03/05&1.93/10& 57& 57\\
 92A&               NGC1380&              SA0& 0&0.006261& 4.80x2.30&  7&-18.50/14&-18.47/12&-18.18/11&0.00/02&1.47/05& 62& 71\\
 94D&               NGC4526&        SA(s)b:sp& 3&0.008427& 4.50x0.80&113&-19.09/12&-19.03/11&-18.79/10&0.00/02&1.32/05& 12& 16\\
98bu&               NGC3368&      SAB(rs)abSy& 2&0.002992& 7.60x5.20&  5&-19.47/16&-19.43/14&-19.18/13&0.33/03&1.01/05& 55& 55\\
\tablecomments{Notation: 
$\rm z^{helio}$ - heliocentric redshift; 
a,b - semi-major and semi-minor axis, arcmin; 
P.A. - position angle, deg; 
$\rm M_B, M_V, M_I$ - reddening corrected SN absolute peak magnitudes in B, V and I-bands, respectively; 
E(B-V) - total SN reddening, as determined by Phillips et al. (1999); 
$\rm \Delta m_{15}$ - decline rate, mag; 
$\sigma$ - $1\sigma$ observational error for the peak magnitudes, the reddening and decline rate, in hundredth of a magnitude; 
$\rm R_p$ - projected angular galactocentric distance, in arcsec; 
$\rm R_{dp}$ - deprojected galactocentric distance for spirals and S0s.}
\enddata
\end{deluxetable}

%% file: tab2.tex
\begin{deluxetable}{cccc}
\tablenum{2}
\tablewidth{0pt}
\tablecaption{Comparison between the parameters of the SNe in 
different types of host galaxies. The significance level P, \% 
from the Kolmogorov-Smirnov test is given (see Section 2 for 
details).\label{tbl-2}}
\tablehead{
\multicolumn{1}{c}{Parameter}& 
\multicolumn{1}{c}{$\rm N_{Ell+S0}$}&
\multicolumn{1}{c}{$\rm N_{Spiral}$}&
\multicolumn{1}{c}{P,\%}}
\startdata
$\rm M_B$&17&35&2\\
$\rm M_V$&17&35&2\\
$\rm M_I$&14&32&7\\
$\rm (B-V)_0$&21&38&90\\
$\rm \Delta m_{15}$&23&39&0.001\\
$\rm E(B-V)$&23&39&12\\
\enddata
\end{deluxetable}

%% file: tab3.tex
\begin{deluxetable}{ccrrrr}
\tabletypesize{\scriptsize}
\tablenum{3}
\tablewidth{0pt}
\tablecaption{Fit coefficients to a linear relation Y=A+B*X for the 
radial gradients of the absolute magnitudes, residuals from the peak 
brightness vs. decline rate relation, and the colors of the SNe Ia 
from our sample.\label{tbl-3}}
\tablehead{
\multicolumn{1}{c}{X}& \multicolumn{1}{c}{Y}& \multicolumn{1}{c}{\#}&
\multicolumn{1}{c}{A}& \multicolumn{1}{c}{B}& \multicolumn{1}{c}{$\sigma$}
}
\startdata
\multicolumn{6}{c}{All Types of Host Galaxies}\\
$\rm R_{dp}^{Kpc}$&          $\rm M_B$&56&-19.419(0.044)& 0.022(0.006)&0.350\\
$\rm R_{dp}^{Kpc}$&     $\rm \Delta_B$&56& -0.060(0.065)& 0.004(0.008)&0.210\\
$\rm R_{dp}/a$    &          $\rm M_B$&52&-19.368(0.044)& 0.245(0.087)&0.340\\
$\rm R_{dp}/a$    &     $\rm \Delta_B$&52& -0.065(0.067)& 0.100(0.129)&0.211\\
$\rm R_{dp}^{Kpc}$&$\rm \Delta m_{15}$&56&  1.055(0.013)& 0.021(0.002)&0.216\\
$\rm R_{dp}/a$    &$\rm \Delta m_{15}$&52&  1.094(0.014)& 0.201(0.024)&0.235\\
$\rm R_{dp}^{Kpc}$&      $\rm (B-V)_0$&56& -0.039(0.058)& 0.001(0.007)&0.095\\
$\rm R_{dp}/a$    &      $\rm (B-V)_0$&52& -0.033(0.058)&-0.003(0.114)&0.097\\
$\rm R_{dp}^{Kpc}$&      $\rm (V-I)_0$&50& -0.310(0.056)& 0.003(0.007)&1.907\\
$\rm R_{dp}/a$    &      $\rm (V-I)_0$&46& -0.296(0.058)&-0.003(0.103)&1.989\\
\multicolumn{6}{c}{Elliptical and S0 Host Galaxies}\\
$\rm R_{dp}^{Kpc}$&          $\rm M_B$&20&-19.153(0.081)& 0.006(0.008)&0.376\\
$\rm R_{dp}^{Kpc}$&     $\rm \Delta_B$&20& -0.010(0.121)& 0.000(0.011)&0.209\\
$\rm R_{dp}/a$    &          $\rm M_B$&19&-19.138(0.074)& 0.091(0.103)&0.379\\
$\rm R_{dp}/a$    &     $\rm \Delta_B$&19& -0.053(0.114)& 0.069(0.152)&0.214\\
$\rm R_{dp}^{Kpc}$&$\rm \Delta m_{15}$&20&  1.279(0.024)& 0.010(0.002)&0.211\\
$\rm R_{dp}/a$    &$\rm \Delta m_{15}$&19&  1.336(0.025)& 0.058(0.029)&0.218\\
$\rm R_{dp}^{Kpc}$&      $\rm (B-V)_0$&20& -0.029(0.106)&-0.001(0.010)&0.119\\
$\rm R_{dp}/a$    &      $\rm (B-V)_0$&19& -0.021(0.096)&-0.019(0.134)&0.120\\
$\rm R_{dp}^{Kpc}$&      $\rm (V-I)_0$&17& -0.382(0.119)& 0.007(0.010)&0.138\\
$\rm R_{dp}/a$    &      $\rm (V-I)_0$&16& -0.345(0.102)& 0.043(0.130)&0.117\\
\multicolumn{6}{c}{Spiral Host Galaxies}\\
$\rm R_{dp}^{Kpc}$&          $\rm M_B$&36&-19.495(0.064)& 0.022(0.015)&0.268\\
$\rm R_{dp}^{Kpc}$&     $\rm \Delta_B$&36& -0.113(0.098)& 0.016(0.021)&0.212\\
$\rm R_{dp}/a$    &          $\rm M_B$&33&-19.411(0.073)& 0.032(0.245)&0.274\\
$\rm R_{dp}/a$    &     $\rm \Delta_B$&33& -0.085(0.103)& 0.012(0.022)&0.214\\
$\rm R_{dp}^{Kpc}$&$\rm \Delta m_{15}$&36&  1.003(0.020)& 0.016(0.005)&0.182\\
$\rm R_{dp}/a$    &$\rm \Delta m_{15}$&33&  1.116(0.024)&-0.258(0.089)&0.187\\
$\rm R_{dp}^{Kpc}$&      $\rm (B-V)_0$&36& -0.057(0.086)& 0.006(0.019)&0.084\\
$\rm R_{dp}/a$    &      $\rm (B-V)_0$&33& -0.051(0.098)& 0.057(0.327)&0.082\\
$\rm R_{dp}^{Kpc}$&      $\rm (V-I)_0$&33& -0.303(0.081)& 0.005(0.019)&2.357\\
$\rm R_{dp}/a$    &      $\rm (V-I)_0$&30& -0.268(0.094)&-0.061(0.316)&2.474\\
\tablecomments{Notation: 
$\rm R_{dp}^{Kpc}$ - the projected galactocentric distance of the SN, Kpc;  
$\rm R_{dp}/a$ - the projected galactocentric distance of the SN normalized 
	to the semi-major axis of the host galaxy; 
$\rm M_B$ - the reddening corrected absolute peak B-band magnitude of the SN; 
$\Delta_B$ - the residual from the SN absolute B-band peak luminosity vs. 
	the decline rate relation, mag; 
$\sigma$ - standard deviation, mag.}
\enddata
\end{deluxetable}